\documentclass[a4paper,11pt]{article}

 \usepackage{amsfonts,amsthm,amssymb}
 \usepackage{latexsym,color, epsf}

 \def\picill#1by#2(#3)
 {\vbox to #2 {\hrule width #1 height 0pt depth 0pt
 \vfill\epsffile{#3}}}

\newcommand{\eq}{\begin{equation}}
\newcommand{\en}{\end{equation}}
\newcommand{\eqa}{\begin{eqnarray}}
\newcommand{\ena}{\end{eqnarray}}

\begin{document}

\setlength{\unitlength}{1mm}

\thispagestyle{empty}

%\begin{flushright}
%Draft for Internal Circulations\\
%v1: 4/18/06; v2: 7/30/06; v3: 11/18/06: v4: 3/12/07; v5: 3/20/07; v6: 3/26/07
%\end{flushright}

%\vspace*{0.1cm}

 \begin{center}
  {\bf \small Topological-Like Features in Diagrammatical Quantum
 Circuits } \footnote{This manuscript is a formal written version
 of Y. Zhang's talk at the workshop ``Cats, Kets and Cloisters"
 (Computing Laboratory, Oxford University, July 17-23, 2006), in
 which he has been proposing categorical quantum physics \& information
 to be described by dagger ribbon categories and emphasizing the functor
 between Abramsky and Coecke's categorical quantum mechanics and his
 extended Temperley--Lieb categorical approach to be the same type as
 those defining topological quantum field theories. As a theoretical
 physicist, however, the proposer
 himself has to admit that these arguments are rather mathematical type
 so that they are hardly appreciated by physicists because physics is such
 a great field including various kinds of topics and topology only plays
 important roles in a limited number of physical problems in the present
 knowledge. On the other hand, this proposal is suggesting either that
 fundamental objects in the physical world are string-like (even brane-like)
 and satisfy the braid statistics or that quasi-particles of many-body systems
 (or fundamental particles at the Planck energy scale) obey the braid statistics
 and have an effective (or a new internal) degree of freedom called the
 ``twist spin",
 so that the braiding and twist operations for defining ribbon categories would
 obtain a reasonable and physical interpretation. Furthermore, this name
 ``categorical quantum physics and information" hereby refers to quantum physics
 and information which can be recast in terms of the language of categories,
 and it is a simple and intuitional generalization of the name ``categorical
 quantum mechanics" because the latter does not recognize conformal field
 theories, topological quantum field theories, quantum gravity and string
 theories which have been already described in the categorical framework by
 different research groups. Moreover,  the proposal {\em categorical quantum
 physics and information} has been strongly motivated by the present study in
 quantum information phenomena and theory, and it is aimed at setting up a
 theoretical platform on which both categorical quantum mechanics and topological
 quantum computing by Freedman, Larsen and Wang are allowed to stand.
    }

  \vspace{.2cm}

 Yong Zhang${}^{a}$
 and Louis H. Kauffman${}^{b}$ \footnote{yong@physics.utah.edu; kauffman@uic.edu} \\[.2cm]

${}^a$ Department of Physics, University of Utah \\
  115 S, 1400 E, Room 201, Salt Lake City, UT 84112-0830
\\[0.1cm]

 ${}^b$  Department of Mathematics, Statistics and Computer Science\\
 University of Illinois at Chicago, 851 South Morgan Street\\
 Chicago, IL, 60607-7045, USA \\[0.1cm]

\end{center}

\vspace{0.2cm}

\begin{center}
\parbox{14cm}{
\centerline{\small  \bf Abstract}  \noindent\\

In this paper, we revisit topological-like features in the extended
Temperley--Lieb diagrammatical representation for quantum circuits
including the teleportation, dense coding and entanglement swapping.
We perform these quantum circuits and derive characteristic
equations for them with the help of topological-like operations.
Furthermore, we comment on known diagrammatical approaches to
quantum information phenomena from the perspectives of both tensor
categories and topological quantum field theories. Moreover, we
remark on the proposal for categorical quantum physics and
information to be described by dagger ribbon categories.

}

\end{center}

\vspace{.2cm}

\begin{tabbing}

\end{tabbing}

%---------------------------------------------------------
\newpage

\section{Introduction}

It is well known that diagrams are capable of catching essential
points from the global view so that they can express complicated
algebraic objects in a much simpler style. Recently, there have been
several diagrammatical approaches proposed to study quantum
information phenomena \cite{nielsen}.  Abramsky and Coecke
\cite{coecke1, coecke2,coecke3} exploit a generalized diagrammatical
representation for tensor categories as a substantial extension of
Dirac's notation to describe quantum information protocols in the
language of strongly compact closed categories. Kauffman and
Lomonaco \cite{kauffman1, kauffman2} show the relationship between
the teleportation procedure and the diagrammatical matrix formalism
used in quantum topology and they call it the teleportation
topology. Griffiths et al. \cite{griffths} devise a set of atemporal
diagrams without reference to time to present quantum circuits.

Furthermore, instead of strongly compact closed categories
\cite{coecke1, coecke2,coecke3}, Zhang \cite{yong1,yong2} proposes
that the Temperley--Lieb (TL) algebra \cite{lieb} under local
unitary transformations underlies quantum information protocols
involving maximally entangled states, projective measurements and
local unitary transformations, and he names the extended TL category
for a collection of all the TL algebras under local unitary
transformations. Also, various descriptions for the quantum
teleportation are found to have a unified description in terms of
the extended TL configurations. See \cite{yong1,yong2} for this
proposal and consult \cite{ZKW,ZKW1} for an introduction to the TL
algebra and the Brauer algebra (i.e., the TL algebra with
permutation) \cite{brauer}. Moreover, Kauffman and Lomonaco
\cite{kauffman3} relate the $0$-dimensional cobordism category to
the Dirac notation of bras and kets and to the quantum
teleportation.

In this paper, we go further to explore
 ``topological" \footnote{In this paper, we only have ``topological" or
 topological-like deformations because topological deformations of the
 diagrams are only allowed if they do not change the algebraic interpretation
 or if they correspond to an algebra identity. Our descriptions for
 quantum circuits are also ``topological" or topological-like because
 we are showing that some diagrammatical representations of quantum circuits
 are subject to limited deformations where the representation of the algebra
 by diagrams corresponds to such deformations.}
 features in the extended TL diagrammatical representation for a quantum
circuit. We define topological-like operations as continuous
deformations of a diagrammatical configuration, and with the help of
them, perform quantum circuits and derive characteristic equations
for the teleportation \cite{bennett,vaidman,eriz,braunstein}, dense
coding and entanglement swapping \cite{ekert}. Besides this, we will
contribute a section for comments on known diagrammatical approaches
to quantum information phenomena including  diagrammatics for
categorical quantum mechanics \cite{coecke1,coecke2,coecke3},
atemporal diagrammatical representation \cite{griffths}, the
extended TL categorical approach \cite{yong1,yong2} and
0-dimensional cobordism category \cite{kauffman3}, and then explain
the suggestion (in the first footnote) that all of them are related
to diagrammatics for tensor categories and the mapping between
categorical quantum mechanics and the extended TL categorical
approach (i.e., 0-dimensional cobordism category or the Brauer
category) is the same type of functor as those defining topological
quantum field theories (TQFT) \cite{atiyah}. Eventually, we make
reliable reasons for the proposal (in the first footnote) that
dagger ribbon categories are responsible for the description of
categorical quantum physics \& information.

The plan of this paper is organized as follows. The extended TL
diagrammatical rules are revisited Section 2, examples for
``topological" operations are listed Section 3, the teleportation
\cite{vaidman, eriz} and entanglement swapping \cite{ekert} are
performed via ``topological" operations Section 4, and
characteristic equations for the teleportation, dense coding and
entanglement swapping are respectively derived Section 5. Comments
on known diagrammatical approaches are made Section 6. Concluding
remarks are on observations for categorical quantum physics \&
information. A brief introduction to the extended TL algebra and
various categorical structures exploited in this manuscript are
respectively made Appendix A and Appendix B.

 \section{Extended Temperley--Lieb diagrammatical rules}

 Maximally entangled states with interesting algebraic properties
 play key roles in quantum information and computation. We review
 extended TL diagrammatical rules \cite{yong1,yong2} for mapping
 every diagrammatical element to an algebraic term in order to
 describe algebraic objects in terms of maximally entangled states.

 \subsection{Notations for maximally entangled states}

 The vectors $|e_i\rangle$, $i=0,1,\cdots d-1$ form a set of orthonormal
 bases for a  $d$-dimension Hilbert space $\cal H$, and the covectors
 $\langle e_i|$ are chosen for its dual Hilbert space
 ${\cal H}^\dag$, \eq
 \sum_{i=0}^{d-1} |e_i\rangle \langle e_i |=1\!\! 1_d, \qquad
 \langle e_j| e_i\rangle =\delta_{ij}, \qquad i,j=0,1,\cdots d-1,
\en where $\delta_{ij}$ is the Kronecker symbol and $1\!\! 1_d$
denotes the $d \times d$ unit matrix. A maximally bipartite
entangled state vector $|\Omega\rangle$ and its dual state vector
$\langle \Omega|$ have the forms, \eq
 |\Omega\rangle=\frac 1 {\sqrt{d}} \sum_{i=0}^{d-1} |e_i\otimes
 e_i\rangle,
 \qquad \langle \Omega | = \frac 1 {\sqrt{d}} \sum_{i=0}^{d-1}
 \langle e_i\otimes e_i |, \en

The action of a bounded linear operator $M$ in the Hilbert space
$\cal H$ on $|\Omega\rangle$ satisfies \eq \label{matrix} (M\otimes
1\!\! 1_d) |\Omega\rangle
  =(1\!\! 1_d \otimes M^T) |\Omega\rangle,  \qquad M^T_{ij}=M_{ji},
  \qquad M_{ij}=\langle e_i | M |e_j \rangle, \en
where the upper index $T$ denotes the transpose, and hence this is
permitted to move the local action of the operator $M$ from the
Hilbert space to the other Hilbert space as it acts on
$|\Omega\rangle$. The trace of two operators $M$ and $N$ can be
represented by an inner product between maximally entangled state
vectors,
 \eq
  tr(M N)=d\cdot\langle \Omega | (M \otimes 1\!\! 1_d)
 (N \otimes 1\!\! 1_d) |\Omega\rangle.
\en The transfer operator $T_{BC}$, sending a quantum state from
Charlie to Bob,  is recognized to be another inner product between
${}_{CA}\langle \Omega|$ and $|\Omega\rangle_{AB}$,
 \eq \label{transfer}
 T_{BC}\equiv\sum_{i=0}^{d-1} |e_i\rangle_B\,\, {}_C\langle e_i |, \qquad
 T_{BC}|\psi\rangle_C = |\psi\rangle_B, \qquad
 T_{BC}=d\cdot {}_{CA}\langle \Omega | \Omega\rangle_{AB},
\en which has been exploited by Braunstein et al. in the
mathematical description for quantum teleportation schemes
\cite{braunstein}.

 The maximally entangled vector $|\Omega_n\rangle$ is a local unitary
 transformation of $|\Omega\rangle$, i.e.,
 $|\Omega_n\rangle =(U_n \otimes 1\!\! 1_d)  |\Omega\rangle$,
  and the set of unitary operators $U_n$ satisfies the
  orthogonal relation $tr(U_n^\dag U_m)=d\, \delta_{nm}$,
  which leads to the following properties, \eq
 \langle \Omega_n |\Omega_m \rangle =\delta_{nm}, \qquad \sum_{n=1}^{d^2}
  |\Omega_n\rangle \langle \Omega_n|=1\!\! 1_d,
  \qquad n,m=1,\cdots d^2,\en
  where the upper index $\dag$ denotes the adjoint.
Introduce the symbol $\omega_n$ for the maximally entangled state
$|\Omega_n\rangle \langle \Omega_n|$ and especially denote
$|\Omega\rangle\langle \Omega|$ by $\omega$, namely,  \eq
 \omega\equiv|\Omega\rangle \langle \Omega|,
 \qquad \omega_n\equiv|\Omega_n\rangle \langle
 \Omega_n|, \qquad U_1=1\!\! 1_d,
\en and the set of $\omega_n$, $n=1,2,\cdots d^2$ forms a set of
 observables over an output parameter space.

\subsection{Extended TL diagrammatical rules}

\begin{figure}
\begin{center}
\epsfxsize=13.cm \epsffile{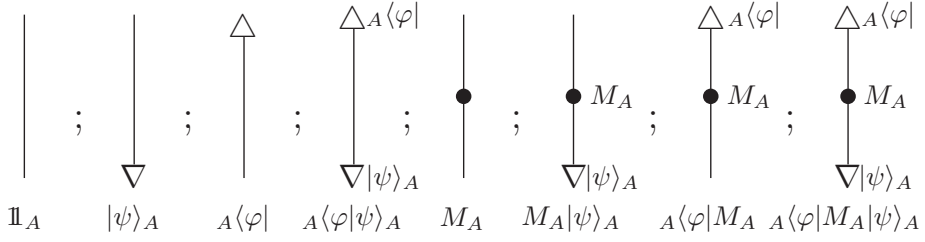} \caption{Straight lines
without or with points.} \label{fig1}
\end{center}
\end{figure}

Three pieces of extended TL diagrammatical rules are devised for
assigning a diagram to a given algebraic object. The first is our
convention; the second explains what straight lines and oblique
lines represent; the third describes various configurations
 in terms of cups and caps.

{\bf Rule 1}. Read an algebraic object such as an inner product from
the left-hand side to the right-hand side and draw a  diagram from
the top to the bottom. Represent the operator $M$ by a solid point,
its adjoint operator $M^\dag$ by a small circle, its transposed
operator $M^T$ by a solid point with a cross line and its complex
conjugation operator $M^\ast$ by a small circle with a cross line.
Denote the Dirac ket by the symbol $\nabla$ and the Dirac bra by the
symbol $\triangle$.

\begin{figure}
\begin{center}
\epsfxsize=13.cm \epsffile{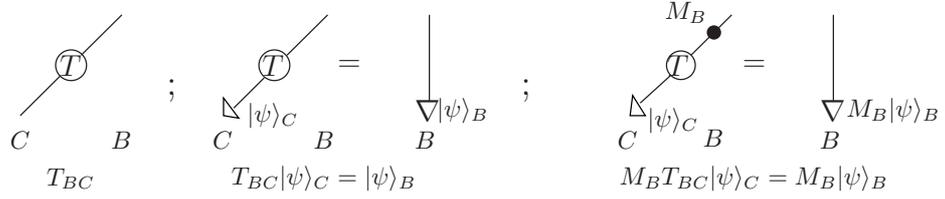} \caption{Oblique line for
transfer operator.} \label{fig2}
\end{center}
\end{figure}

{\bf Rule 2}. See Figure 1. A straight line of type $A$ denotes the
identity operator $1\!\! 1_A$ for the system $A$. Straight lines of
type $A$ with a bottom $\nabla$ or top $\triangle$ describe a vector
$|\psi\rangle_A$, covector ${}_A\langle \varphi|$, and an inner
product ${}_A\langle \varphi|\psi\rangle_A$ in the system $A$,
respectively. Straight lines of type $A$ with a middle solid point
or bottom $\nabla$ or
 top $\triangle$  describe an operator $M_A$, a vector $M_A|\psi\rangle_A$,
 a covector ${}_A\langle \varphi| M_A$,  and an inner
 product ${}_A\langle \varphi | M_A |\psi\rangle_A$, respectively.

See Figure 2. An oblique line from the system $C$ to the system $B$
describes the transfer operator $T_{BC}$, and its solid point or
bottom $\nabla$ or top $\triangle$  have the same interpretations as
those on a straight line of type $A$ in Figure 1.

\begin{figure}
\begin{center}
\epsfxsize=13.cm \epsffile{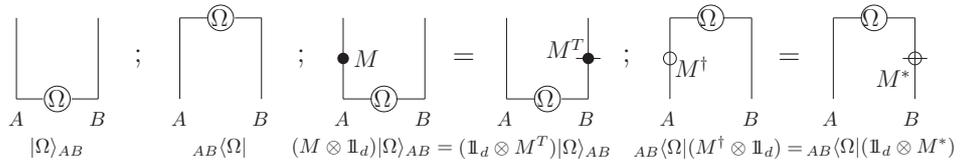} \caption{Cups and caps without
or with points.} \label{fig3}
\end{center}
\end{figure}

{\bf Rule 3}. See Figure 3. A cup denotes the maximally bipartite
entangled state vector $|\Omega\rangle$ and a cap does for its dual
$\langle \Omega|$. A cup with a middle solid point on its one branch
describes a local action of the operator $M$ on $|\Omega\rangle$,
and this solid point can flow to its other branch and is replaced by
a solid point with a cross line representing $M^T$. The same happens
for a cap except that a solid point is replaced by a small circle to
distinguish the operator $M$ from its adjoint operator $M^\dag$.

A cup and a cap can form different sorts of configurations. See
Figure 4. As a cup is at the top and a cap is at the bottom  for the
same composite system, this configuration is assigned to the
projector $|\Omega\rangle \langle \Omega |$. As a cap is at the top
and a cup is at the bottom for the same composite system, this
diagram describes an inner product $\langle \Omega|\Omega\rangle=1$
by a closed circle. As a cup is at the bottom for the composite
system ${\cal H}_C \otimes {\cal H}_A$ and a cap is at the top for
the composite system ${\cal H}_A\otimes {\cal H}_B$, that is an
oblique line representing the transfer operator $T_{BC}$ with the
normalization factor $\frac 1 d$.

\begin{figure}
\begin{center}
\epsfxsize=13.5cm \epsffile{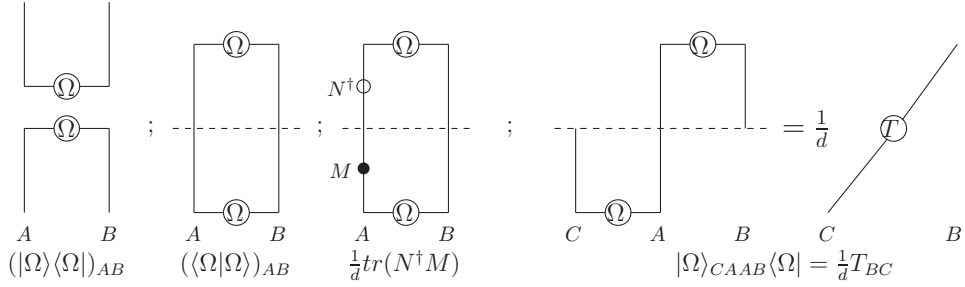} \caption{Three kinds of
combinations of a cup and a cap.} \label{fig4}
\end{center}
\end{figure}

Additionally, as a cup has a local action of the operator $M$ and a
cap has a local action of the operator $N^\dag$, the resulted circle
with a solid point for $M$ and a small circle for $N^\dag$
represents the trace $\frac 1 d tr (M N^\dag)$. As conventions, we
describe a trace of operators by a closed circle with solid points
or small circles, and assign each cap or cup a normalization
 factor $\frac 1 {\sqrt d}$ and a circle a normalization factor $d$.

Note that cups and caps are well known configurations in knot theory
and statistics mechanics. They were used by Wu \cite{wu} in
statistical mechanics, and exploited by Kauffman \cite{kauffman4}
for diagrammatically representing the Temperely-Lieb algebra soon
after Jones's work \cite{jones}. These configurations  are nowadays
 called Brauer diagrams \cite{brauer} or Kauffman diagrams \cite{kauffman4}.

\section{Examples for ``Topological" operations}

Topological-like operations are defined as continuous deformations
of diagrammatical configurations, and three typical examples
exploited in what follows are presented.

\begin{figure}
\begin{center}
\epsfxsize=11.cm \epsffile{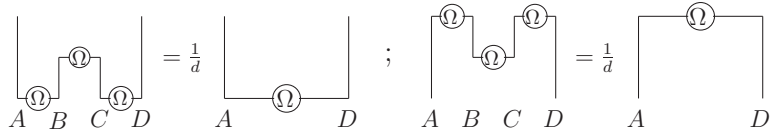} \caption{Cup and cap via
compositions of cups and caps. } \label{fig5}
\end{center}
\end{figure}

  See Figure 5 for two kinds of compositions of cups and caps.
  The configuration of a cup (cap) can be regarded as
  a composition of a series of cups and caps. The cup state $|\Omega\rangle_{AD}$
 is obtained by connecting the cap state ${}_{BC}\langle \Omega|$ with the
 cup states $|\Omega\rangle_{AB}$ and $|\Omega\rangle_{CD}$,
 namely proved  by
 \eq
 ({}_{BC}\langle \Omega|)(|\Omega\rangle_{AB})(|\Omega\rangle_{CD})
  =\frac 1 d |\Omega\rangle_{AD},
 \en
and the cap state ${}_{AD}\langle \Omega|$ is a composition of the
cap states ${}_{AB}\langle \Omega|$, ${}_{CD}\langle \Omega|$ and
the cup state $|\Omega\rangle_{BC}$, specified by
 \eq
({}_{AB}\langle \Omega|)({}_{CD}\langle
\Omega|)(|\Omega\rangle_{BC})
 =\frac 1 d\,\, {}_{AD} \langle \Omega|.
 \en

\begin{figure}
\begin{center}
\epsfxsize=11.cm \epsffile{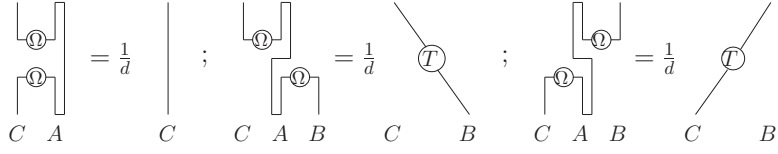} \caption{Three kinds of
partial traces between a cup and a cap.} \label{fig6}
\end{center}
\end{figure}

 See Figure 6 for two sorts of diagrammatical partial traces.
 The partial trace of a composite system denotes the
 summation over its subsystem, for example,
 \eq tr_{A}(|e^C_i\otimes e^A_j\rangle\langle e^A_l\otimes e^B_m|)
 =|e^C_i\rangle \langle e^B_m| \delta_{jl},\qquad
 tr_A(|e^A_j\rangle\langle e^A_l |)=\delta_{jl}, \en
 while the trace is defined as the summation over the entire composite system,
 \eq
 tr_{CA} (|e^C_i\otimes e^A_j\rangle\langle e^C_l\otimes
 e^A_m|)=\delta_{il}\delta_{jm}, \qquad i,j,l,m=0,1,\cdots d-1.
 \en
 The first type of diagrammatical partial trace leads to a straight
 line, verified by
 \eq
 tr_A(|\Omega\rangle_{CA}\, {}_{CA}\langle \Omega| )=\frac 1 d
 \sum_{i,j=0}^{d-1} tr_A (|e^C_i\otimes e^A_i\rangle \langle e^C_j \otimes e^A_j|)
 =\frac 1 d (1\!\! 1_d)_C,
 \en
and the second type of diagrammatical partial traces yield oblique
lines for the transfer operators $T_{CB}$ and $T_{BC}$, which are
algebraically represented by
 \eq
 \frac 1 d T_{CB}=tr_A(|\Omega\rangle_{CA}\, {}_{AB}\langle \Omega|
 ), \qquad
 \frac 1 d T_{BC}=tr_A({}_{CA}\langle \Omega |\Omega\rangle_{AB}).
 \en

\begin{figure}
\begin{center}
\epsfxsize=11.cm \epsffile{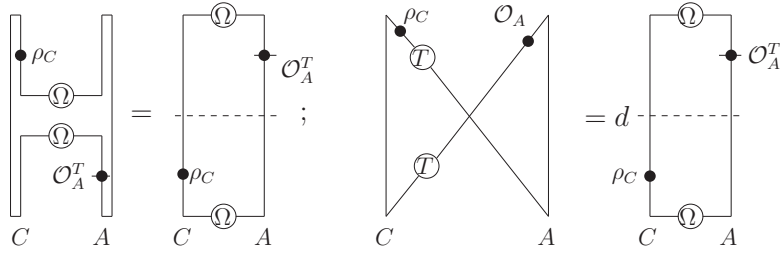} \caption{Closed circles via
cup and cap, and  oblique lines} \label{fig7}
\end{center}
\end{figure}

See Figure 7 for how to form closed circles in three distinct ways.
A top cup with a bottom cap forms the same closed circle as a top
cap and a bottom cup, as is revealed in the algebraic expression,
\eq
 tr_{CA}( (\rho_C\otimes 1\!\! 1_d |\Omega\rangle_{CA})
 ({}_{CA}\langle \Omega| 1\!\! 1_d\otimes {\cal O}_A^T))
 = {}_{CA}\langle \Omega| (1\!\! 1_d\otimes {\cal O}_A^T)(\rho_C\otimes
1\!\! 1_d) |\Omega\rangle_{CA}, \en
 where $\rho_C$ and ${\cal O}_A$ are bounded linear operators in the
 $d$-dimensional Hilbert space. A closed circle formed by two oblique
 lines denotes the same trace as a top cap with a bottom cup,
 as can be algebraically proved,
 \eq
 \label{oblique}
 tr_{CA} ((\rho_C T_{CA})({\cal O}_A T_{AC}))
  = d\,\cdot {}_{CA}\langle \Omega| (\rho_C\otimes 1\!\! 1_d) (1\!\!
  1_d\otimes {\cal O}^T_A) |\Omega\rangle_{CA}.
 \en

In the extended TL diagrammatical representation for a quantum
circuit, we perform its presumed functions using topological-like
operations defined as above.

 \section{``Topological" descriptions for quantum circuits}

We study topological-like descriptions for the teleportation and
entanglement swapping which are stimulating examples for extended TL
diagrammatical quantum circuits.

 \subsection{``Topological"  description for teleportation}

 Teleportation \cite{bennett} can be observed from the viewpoint of
 quantum measurement \cite{vaidman,eriz}. The maximally entangled
 state $|\Omega\rangle_{AB}$ shared by Alice and Bob is created in
 the quantum measurement denoted by the projector $(|\Omega\rangle
 \langle \Omega|)_{AB}$, while the quantum measurement performed by
 Alice in the composite system of Charlie and herself is represented
 by the projector $(|\Omega_n\rangle \langle \Omega_n|)_{CA}$.
 Therefore the teleportation equation has the following formulation,
 \eq
 \label{mtele}
 (|\Omega_n\rangle \langle \Omega_n|
 \otimes 1\!\! 1_d) (|\psi\rangle \otimes |\Omega \rangle \langle \Omega| )
 =\frac 1 d (|\Omega_n\rangle \otimes 1\!\! 1_d )
  (1\!\! 1_d \otimes (1\!\! 1_d \otimes U_n^\dag
  |\psi\rangle) \langle \Omega| ),  \en where the lower indices $A, B, C$ are
  omitted for convenience and there are $d^2$ distinguished
  classical channels between Alice and Bob due to $n=1,\cdots, d^2$.

  \begin{figure}
  \begin{center}
  \epsfxsize=8.8cm \epsffile{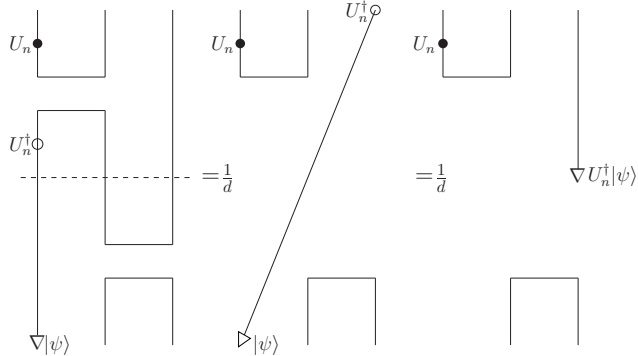}
  \caption{``Topological" description for the teleportation.} \label{fig8}
  \end{center}
  \end{figure}

  We make a ``topological" description for the teleportation based on
  quantum measurements. It is encoded in Figure 8 where the symbols $\Omega$
  and $T$ are omitted for simplicity, by reading the teleportation equation
  (\ref{mtele}) from the left hand-side to the right hand-side and drawing
  Figure 8 from the top to the bottom in view of extended TL diagrammatical rules.
  Move the unitary operator $U_n^\dag$ from Charlie's system to Bob's system
  along the path formed by a top cap and a bottom cup; apply the ``topological"
  operation by straightening the configuration of the top cap and bottom cup
  into an oblique line; transport a unknown quantum state $|\psi\rangle_C$ along
  the oblique line from Charlie to Bob. Finally, Charlie has a quantum state
  $U_n^\dag |\psi\rangle_B$ and applies the local unitary transformation $U_n$
  to obtain $|\psi\rangle_B$. Note that in the extended TL diagrammatical recipe for
  the teleportation, the cup and the cap do not straighten to the identity
  but rather to a unitary transformation that depends upon the measurement outcome.

  Furthermore, Figure 8 is presenting a suitable diagrammatical framework for
  unifying various sorts of descriptions for teleportation, see \cite{yong1,yong2}
  for the detail. It can include Vaidman's continuous teleportation \cite{vaidman,eriz}
  and other discrete teleportation schemes. Its enclosure (Figure 10) represents
  Werner's tight teleportation scheme \cite{werner}. Removing ``irrelevant"
  parts (the top cup with the solid point for $U_n$ and the bottom cap)
  leads to the well known configuration for the quantum information flow in the
  literature \cite{coecke1,coecke2,coecke3,kauffman1,kauffman2,kauffman3,griffths}.

 Moreover, a cup-over-cap represents a projector which formally denotes
 a quantum measurement and has little to do with how states are
 actually prepared and measured in the laboratory. It is a generator
 of the TL algebra and hence Figure 8 is typical configuration of
 the extended TL algebra, i.e., the TL algebra under local unitary
 transformations, see Appendix A. Hence a natural connection between
 quantum information and the TL algebra has been recognized
 this way \cite{yong1,yong2}.

  \subsection{``Topological" description for entanglement swapping}

 Entanglement swapping \cite{ekert} produces the entanglement between
 two independent systems as a consequence of quantum measurements
 instead of physical interactions. Alice has a maximally entangled
 bipartite state $|\Omega_l\rangle^{A}_{ab}$ for particles $a, b$ and Bob
 has $|\Omega_m\rangle^B_{cd}$ for particles $c,d$. They are independently
 created and do not share common history. Alice applies a quantum measurement
 denoted by
 $1\!\! 1_d\otimes (|\Omega_n\rangle \langle \Omega_n|)_{bc} \otimes 1\!\! 1_d$ to
 the product state of $|\Omega_l\rangle^{A}_{ab}$ and $|\Omega_m\rangle^B_{cd}$
 so that the entanglement swapped state $|\Omega_{lnm}\rangle^{AB}_{ad}$
 is a maximally entangled bipartite state shared by Alice and Bob for
 particles $a,d$, i.e.,
    \eqa
   \label{swap}
  & & (1\!\! 1_d\otimes (|\Omega_n\rangle \langle \Omega_n|)_{bc} \otimes 1\!\! 1_d)
  (|\Omega_l\rangle^{A}_{ab} \otimes |\Omega_m\rangle^B_{cd})
  \nonumber\\
  & &= \frac 1 d ( 1\!\! 1_d\otimes |\Omega_n\rangle_{bc} \otimes 1\!\! 1_d)
  \frac 1 {\sqrt d}\sum_{i=0}^{d-1}
  ( U_l U_n^\ast U_m  |e_i\rangle^A_a\otimes 1\!\! 1_d
  \otimes 1\!\! 1_d\otimes |e_i\rangle^B_d)  \nonumber\\
  & &\equiv \frac 1 d ( 1\!\! 1_d\otimes |\Omega_n\rangle_{bc} \otimes 1\!\! 1_d)\,\,
   |\Omega_{lnm}\rangle^{AB}_{ad}.
    \ena
In other words, the entanglement swapping reduces a four-particle
state $|\Omega_l\rangle^{A}_{ab}\otimes |\Omega_m\rangle^B_{cd}$ to
a bipartite entangled state $|\Omega_{lnm}\rangle^{AB}_{ad}$ using
the entangling quantum measurement.

\begin{figure}
\begin{center}
\epsfxsize=9.cm \epsffile{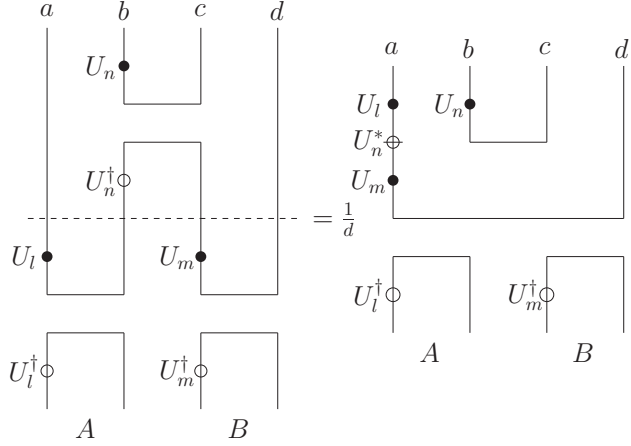}
 \caption{``Topological" description for the entanglement swapping.}
 \label{fig9}
\end{center}
\end{figure}

In the extended TL diagrammatical representation for the
entanglement swapping, Figure 9 in which the omission of symbols
$\Omega$ has no confusion, the entanglement swapping equation
(\ref{swap}) can be proved at the diagrammatical level by collecting
unitary operators $U_l$, $U_n^\ast$ and $U_m$ at the system for the
particle $a$, and then applying ``topological" diagrammatical
operations in Figure 6. Note that Figure 9 is a standard
Temperely--Lieb configuration under local unitary transformations
and the entanglement swapping presents a typical example for the
quantum network consisting of maximally entangled states and local
unitary transformations, see Appendix A or \cite{yong1,yong2}.

 \section{Characteristic equations for quantum circuits}

A characteristic equation for a quantum circuit is derived by
applying a series of topological-like operations to the enclosure of
its extended TL diagrammatical representation. This equation
includes all essential elements for this quantum circuit. We study
characteristic equations for the teleportation, dense coding and
entanglement swapping.

Here all involved finite Hilbert spaces are $d$ dimensional and the
classical channel distinguishes $d^2$ signals, which are called the
tight scheme for quantum information protocols by Werner
\cite{werner}. The density operator $\rho$ has a form
$\rho=|\phi_1\rangle \langle \phi_2 |$.
 The channel $T_n$ describes a local unitary transformation
 $U_n$ on an observable ${\cal O}$, which are given by
 \eq
 T_n ({\cal O} ) =U_n^\dag {\cal O} U_n, \qquad {\cal O} =|\psi_1
 \rangle \langle \psi_2 |, \qquad n=1,2, \cdots d^2.
\en

 \subsection{Characteristic equation for teleportation}

In the tight teleportation scheme \cite{werner}, Charlie has his
density operator $\rho_C=(|\phi_1\rangle \langle \phi_2 |)_C$ to
include a quantum state sent to Bob, while Alice and Bob share the
maximally entangled state $\omega_{AB}=$ $(|\Omega\rangle \langle
\Omega |)_{AB}$. Alice chooses her observables $(\omega_n)_{CA}$ to
make the Bell measurement in the composite system between Charlie
and her and then passes the message labeled by $n$ on to Bob via the
classical channel. Finally, Bob performs a unitary correction on his
observable ${\cal O}_B$ by the quantum channel $T_n$. In terms of
$\rho_C$, $\omega_{AB}$, $(\omega_n)_{CA}$ and $T_n({\cal O}_B)$,
the tight teleportation scheme is summarized in the characteristic
equation,
 \eq
 \label{ttele}
 \sum_{n=1}^{d^2} tr ( (\rho_C\otimes \omega_{AB} )
 ((\omega_n)_{CA} \otimes T_n({\cal O}_B) ))
 =tr(\rho_C {\cal O}_B),
 \en
which catches the crucial point of a successful teleportation, i.e.,
Charlie makes the measurement in his system as he does in Bob's
system although they are independent from each other.

See the left term of Figure 10 where the lower indices $A,B,C$ are
omitted for convenience. The enclosure of its extended TL
diagrammatical representation Figure 8, is obtained by connecting
top boundary points to bottom boundary points in the systems for
Charlie, Alice and Bob, respectively. Working on such the enclosure,
we have two diagrammatical approaches of deriving the characteristic
equation (\ref{ttele}) for the teleportation. The first way is to
move the local unitary operators $U_n^\dag$ and $U_n$ along the
configuration of cups or caps until they meet to yield the identity,
apply ``topological" operations suggested by the second or third
term of Figure 6 and then exploit the right term of Figure 7. The
second way is to combine ``topological" diagrammatical operations
suggested by the first term of Figure 6 with the left term of Figure
7. In addition, we arrange the density operator $\rho$ and
observable $\cal O$ in the same straight line by moving them along
branches of cups or caps.

\begin{figure}
\begin{center}
\epsfxsize=13.5cm \epsffile{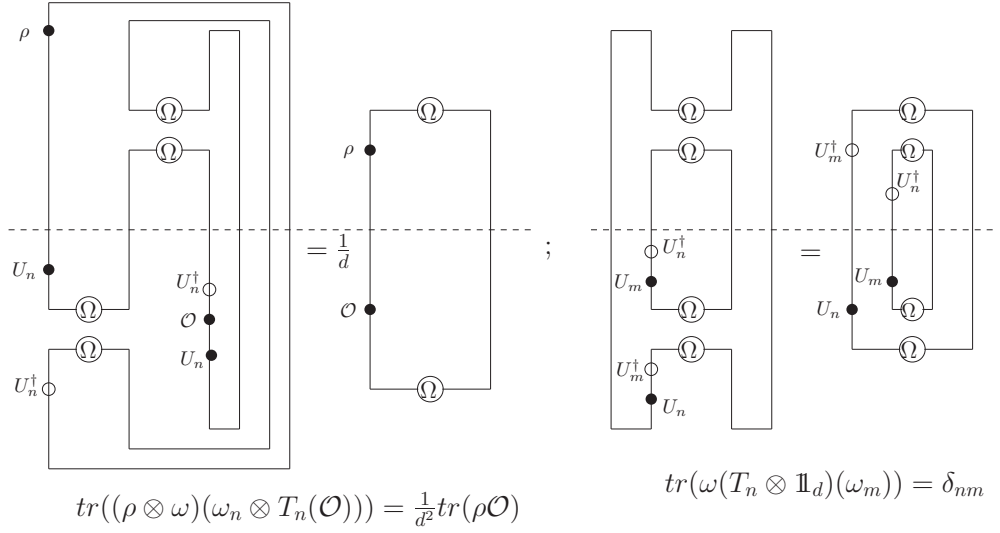} \caption{Characteristic
 equations for teleportation and dense coding.} \label{fig10}
\end{center}
\end{figure}

 \subsection{Characteristic equation for dense coding}

The tight dense coding \cite{werner} can be also performed using
topological-like operations. Alice and Bob share the maximally
entangled state $|\Omega\rangle_{AB}$, and Alice transforms her
state by the channel $T_n$ to encode a message $n$ and then Bob
makes the quantum measurement on an observable $\omega_m$ of his
system. At $n=m$, Bob gets the message. See the right term of Figure
10:  The process of this kind of dense coding is concluded in its
diagrammatical and algebraic characteristic equations, with
``topological" operation by the left term of Figure 7 to be
exploited.

 \subsection{Characteristic equation for entanglement swapping}

\begin{figure}
\begin{center}
\epsfxsize=8.cm \epsffile{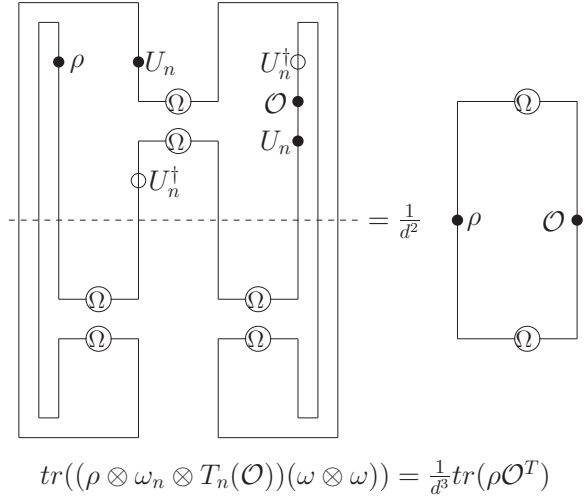} \caption{Characteristic
equation for the entanglement swapping.} \label{fig11}
\end{center}
\end{figure}

 See Figure 11 in which the lower indices $a,b,c,d$ are neglected and
 the transpose ${\cal O}^T$ of the observable ${\cal O}$ is defined by
 \eq {\cal O}^T=\sum_{i,j=0}^{d-1}
 \psi_{1i} \psi_{2j}^\ast |e_j\rangle \langle e_i|,\qquad (\langle
 e_i|)^T\equiv|e_i\rangle. \en
 It is the enclosure of its extended TL diagrammatical representation
  Figure 9. Exploit ``topological" operations by Figure 5 and then that
  by the left term of
 Figure 7 to derive its characteristic equation,
 \eq \label{tightswap}
 \sum_{n=1}^{d^2} tr ((\rho_a\otimes (\omega_n)_{bc} \otimes T_n({\cal O}_d))
 (\omega_{ab}\otimes \omega_{cd})) =\frac 1 d tr (\rho_a {\cal O}_d^T), \en
where the summation is over $n^2$ classical channels. Note that the
entanglement swapping can be used to detect the Bell inequality
\cite{nielsen} although entanglement is yielded via quantum
measurements.

\section{Comments on known diagrammatical approaches}

This section is aimed at commenting on several known diagrammatical
approaches devised for describing quantum information phenomena in
the recent literature by pointing out essential differences and
connections among them. All of them are believed to be various
generalizations of the diagrammatical technique in relation to
tensor categories well-known in the mathematical literature,
stemming from the work by Joyal and Street \cite{joyal} in the
1990's, with important contributions by Turaev and others, (and also
going back to pioneering work by Kelly in the 1970's \cite{kelly1}),
see Kassel' textbook \cite{kassel} and Turaev's book \cite{turaev}
for more relevant references.

Category is a sort of the abstract language to describe a collection
of mathematical objects as well as structure-preserving morphisms
between them. Definitions of various kinds of categories used in the
following have been sketched in Appendix B, they including
categories, monoidal categories (tensor categories), pivotal
categories, dagger pivotal categories, braided monoidal categories,
symmetric monoidal categories, symmetric pivotal categories (compact
closed categories), symmetric dagger pivotal categories (strongly
compact closed categories or 3-tuply categories with duals) and
$d$-dimensional cobordism categories.

\subsection{Categorical quantum mechanics \& information}

Categorical structures for quantum information phenomena can be set
up by regarding physical systems (such as qubits) as objects and
physical operations (such as local unitary transformations) as
morphisms. To propose high-level methods for quantum computation and
information, Abramsky and Coecke \cite{coecke1,coecke2,coecke3}
refine strongly compact categories and exploit them to
comprehensively axiomatize quantum mechanics and study quantum
information protocols, and also make the detailed elaboration of
diagrammatical representation for tensor categories to quantum
mechanics \& information.

Categorical quantum mechanics is a typical example for strongly
compact closed categories, and it has all Hilbert spaces as objects
and linear bounded operators as morphisms. It is equipped with
dagger and dual operations to capture the complex structure of
quantum mechanics where one has transpose and complex conjugation as
separate things with the adjoint distinct from the dual. The
diagrammatical representation for categorical quantum mechanics can
be viewed as a two-dimensional generalization of the Dirac notation.

Strongly compact closed categories are also called 3-tuply
categories with duals by Baez and Dolan \cite{baez1, baez2} or
dagger compact closed categories by Selinger \cite{selinger1}.

\subsection{Atemporal diagrammatical approach}

Griffiths et al. \cite{griffths} devise a system of atemporal
diagrams to describe various elements of a quantum circuit where
``atemporal" means such a diagrammatical representation makes no
reference to time. This kind of diagrammatical representation is
also a sort of generalization of the diagrammatical recipe for
tensor categories. But it allows oriented diagrams with an arrow
from a Hilbert space to its adjoint space, and explicitly reveals
the map-state duality in which a maximally bipartite entangled state
can be identified with a unitary map. Especially, its diagrammatical
prescription on completely positive map, positive operator, super
operator, transition operator and dynamical operator is almost
equivalent to Selinger's CPM construction \cite{selinger1} over
strongly compact closed categories.

In the atemporal diagrammatical approach \cite{griffths}, the
maximally entangled state plays the role of the transposer denoted
by $|{\cal A}\rangle$, a quantum (classical) channel is simply
denoted by lines, and any type of operators can be represented since
lines are allowed to point to any directions. In the extended TL
diagrammatical representation \cite{yong1,yong2}, however, the
maximally entangled state $|\Omega\rangle$ (or $\langle \Omega|$)
plays the kernel role with the cup (or cap) configuration
\footnote{The maximally entangled state $|\Omega\rangle$ and its
adjoint state $\langle\Omega|$ can be respectively  regarded as the
unit and counit mappings defining the pivotal category (see Appendix
B) in which the dual of a morphism is understood to be its ordinary
transposition. This is an important connection between the atemporal
diagrammatical approach \cite{griffths} and the extended TL
categorical approach \cite{yong1,yong2}.}, the quantum channel is a
TL configuration with solid points (or small circles), and labels
for distinct Hilbert spaces are arranged at a horizontal line so
that ``topological" features in diagrammatical quantum circuits can
be explicitly observed.

Moreover. Removing ``irrelevant parts" from Figure 8 leads to the
same quantum channel as Fig. 7(d) \cite{griffths} if one represents
$\Psi$ by a cup and denote $\Phi_j$ by a cap with a small circle for
the local unitary transformation $U_j^\dag$, and as $j=k$ the
teleportation is performed. As one makes a double of Fig. 7(d)
\cite{griffths} with an additional density operator and observable,
he will obtain Werner's tight teleportation scheme \cite{werner},
i.e., the left term of Figure 10.

\subsection{The extended Temperley--Lieb categorical approach}

In this paper together with \cite{yong1, yong2}, motivated by
seeking for topological and algebraic structures underlying various
quantum information phenomena and then obtaining helpful insights
for the application of unitary Yang--Baxter solutions as universal
quantum gates to quantum information and computing, extended TL
diagrammatical rules are set up for quantum information protocols in
terms of maximally entangled states and local unitary
transformations. Topological-like descriptions are made for the
teleportation, dense coding and entanglement swapping using
topological-like operations and characteristic equations are derived
for them.

In \cite{yong1, yong2}, the TL algebra with physical operations
(such as local unitary transformation) is found to present a
suitable mathematical framework for quantum information protocols
including quantum teleportation and entanglement swapping. The TL
category denotes a collection of all TL algebras: the unoriented TL
category is a free dagger pivotal category over self-dual object,
i.e., the oriented TL category is a free dagger pivotal category.
Therefore, the extended TL category represents a collection of all
TL algebras with physical operations. Furthermore,
 the teleportation configuration \cite{yong1,yong2} can be regarded
as the defining configuration for the diagrammatical representation
of the Brauer algebra \cite{brauer} which is the extension of the TL
algebra with a symmetry (permutation or swap or flat crossing)
generator. Similarly, the Brauer category denoting the collection of
all Brauer algebras, has oriented and unoriented generalizations.
Additionally, the Brauer category is a kind of the extended TL
category because a symmetry (swap) can be represented by a linear
combination of the extended TL configurations (see \cite{yong2} for
an example).

Although cup (cap) configurations are exploited in diagrammatical
representations for both categorical quantum mechanics and extended
TL categorical approach, strongly compact closed categories are
symmetric but the TL categories are planar without symmetry. More
significantly, in the description of quantum information protocols
\cite{yong1,yong2} like quantum teleportation and entanglement
swapping, a symmetry operator is not needed and the transposition
$M^T$ of an operator $M$ is found to play the key role\footnote{This
is consistent with the fact that the TL category is a pivotal
category in which no symmetry is imposed and the dual of a morphism
can be identified with its transposition.}.  Moreover, essential
differences between two approaches to the description of quantum
information flow have been presented in detail, see \cite{yong1,
yong2}, where the quantum information flow is created together with
loops, ``irrelevant" cups and caps, and the normalization factor
allowed to be zero.

Note that conceptual differences between Kauffman and Lomonaco's
teleportation topology \cite{kauffman1} and the extended TL
categorical approach are explained clear \cite{yong1, yong2},
together with both approaches appreciating the same matrix
diagrammatical technique.

\subsection{Cob[0]: 0-dimensional cobordism category}

Kauffman and Lomonaco \cite{kauffman2} remark on the 0-dimensional
cobordism category {\bf Cob[0]} with a natural relationship to
quantum mechanics and quantum teleportation, i.e., directly related
to the Dirac notation of bras and kets. The one object of the
 {\bf Cob[0]} category is a single point $p$, i.e., the simplest zero
dimensional manifold, and the other object is the empty set $\ast$
(the empty manifold). A morphism between the point $p$ and another
point $q$ is the line segment with boundary points $p$ and $q$, the
identity morphism to be a map from $p$ to $p$. It is simple to
identity various morphisms between points and the empty set $\ast$
with Dirac bras and kets, together with the scalar product
recognized to be a morphism between $\ast$ and $\ast$. Note that the
{\bf Cob[0]} category is the Brauer category, in other words, the
 {\bf Cob[0]} category without crossings between any of line segments is
the TL category, furthermore, it is also a sort of the extended TL
category.

\subsection{Summarizing comments from the perspective of TQFT}

As a summary of our comments on known diagrammatical approaches, we
know from the first footnote that what is relating categorical
quantum mechanics to the extended TL categorical approach (i.e., the
{\bf Cob[0]} category or Brauer category) is a well known functor
used to define TQFT{\footnote{Also, once we acknowledge that the
functor defining TQFT plays the key role in quantum physics, we may
be suggesting the nature of quantum physics to be determined by the
topology of the ground state of a given physical system, for example
that quantum orders describing distinct phases in the ground state
of many-body systems may be recognized to be topological orders
\cite{levin}.} by Atiyah \cite{atiyah}. He defines a $d$-dimensional
TQFT as a functor from the $d$-dimensional cobordism category {\bf
Cob[d]} to the category {\bf Vect} of all vector spaces and linear
mappings, i.e., a representation of the {\bf Cob[d]}.

Furthermore, a $d$-dimensional TQFT is also a functor from the {\bf
Cob[d]} category to the category {\bf Hilb} of all Hilbert spaces
and linear bounded operators, and this functor can be equipped with
two types of dualities which are exploited to define the transpose,
complex conjugation, and adjoint operation for all morphisms. Here
we call this sort of functor by the strongly compact closed
functor\footnote{According to the proposal in the first footnote
about {\em dagger ribbon categories for categorical quantum physics
and information}, this functor would be better called the {\em
dagger ribbon functor}. } because both the {\bf Cob[d]} category and
{\bf Hilb} category are strongly compact closed categories.

Moreover, it is worthwhile noting \footnote{This note is suggesting
there still remain many fundamental conceptual problems to be solved
and clarified in categorical quantum mechanics (physics), for
example, how to fix the global phase in the superposition principle
of state vectors in a Hilbert space, see \cite{coecke2}.} that,
state preparation and quantum measurements are essential parts of
quantum information and computation, but which can not be directly
described by categorical structures which operate with objects, and
morphisms and functors. This means that in categorical quantum
mechanics we have two conceptual levels: the one is devised for
categorical descriptions; the other is specially designed for state
vectors in the Hilbert space. In other words, in the known
categorical definition of TQFT, state preparation and quantum
measurement have been not considered as seriously as in categorical
quantum mechanics \& information.

\section{Categorical quantum physics \& information}

 In the last section, first of all, ambiguities have to be clarified
 for why the proposal in the first footnote chooses the name
 {\em categorical quantum physics and information} instead of
 categorical quantum mechanics \& information suggested by
 Abramsky and Coecke \cite{coecke1, coecke2,coecke3}. As the unit object
 of pivotal categories (see Appendix B) is identified with the vacuum,
 the unit morphism $\eta_A$ (the counit morphism $\epsilon_A$) can be
 explained as the creation (annihilation) of a pair of a particle and
 its anti-particle from (into) the vacuum with the input (output) of
 enough energy, which is exploiting the language of quantum field theory
 instead of quantum mechanics. Besides this, another explicit point is that
 these categorical structures (as above or see Appendix B) are not just important
 in the formulation of basic quantum mechanics itself (oriented towards
 quantum information) but also for exotic constructions towards TQFT,
 string theories and quantum gravity.

 As we see in this paper, categorical structures have been intensively
 exploited in the study of quantum information and computation.
 Strongly compact closed categories have an example of the category
 {\bf Hilb} including all Hilbert spaces and linear bounded operators
 for recasting axioms of quantum mechanics in the abstract language.
 The extended TL categories are found to be especially fitted for quantum
 information protocols like quantum teleportation and entanglement
 swapping. The ``partially" braided monoidal categories
 \footnote{They are based on \cite{kauffman1,yong3,yong4}
 which study universal quantum computation by combining non-trivial
 unitary braids as  two-qubit universal quantum gates with single qubit
 transformations. ``Partially" means that the naturality condition for
 defining the braided monoidal category can not be satisfied because a
 tensor product of single qubit transformations does not often commute with
 a two-qubit braiding gate.} is suggested to be a mathematical framework for
 describing quantum circuits consisting of single qubit transformations and
 non-trivial unitary braids as universal two-qubit quantum gates. In addition,
 modular tensor categories (see Appendix B) for 2-dimensional TQFT
 (for example, the $SU(2)$ Witten-Chern-Simons theory at roots of unity
 \cite{witten}) are the mathematics framework for topological quantum computing
 by Freedman, Larsen and Wang \cite{flw1, flw2}.

 Besides these applications of various category theories to quantum information
 phenomena, conformal field theories defined by Segal \cite{segal}
 describe processes in string theories as morphisms; loop gravity \cite{smolin}
 has spin network as objects and spin foams as morphisms; a higher-dimensional
 categorical notion called $n$-category is devised by Baez and Dolan
 \cite{baez1,baez2} for reconciling the general relativity and quantum
 mechanics. Especially, a kind of physical interpretation for every element
 of tensor categories has been proposed by Levin and Wen \cite{levin} to
 explain topological orders in condensed matter physics via string-net
 condensation.

 Therefore, we think that we are able to make acceptable reasons for the
 proposal in the first footnote that {\em dagger ribbon  categories}
 (i.e., strongly ribbon categories or ribbon categories with the dagger
 operation, see Appendix B) are a suitable mathematical framework to describe
 {\em categorical quantum physics and information} \footnote{In categorical
 quantum physics and information, it is necessary to introduce state vectors of
 the Hilbert space (which is only an object of the dagger ribbon category) to
 define state preparation and quantum measurement. How to deal with
 classical communication is also  an interesting problem to be discussed in
 the categorical language or quantum approach.}.
 The combination of the left duality with the dagger operation ensures the complex
 structure and complex conjugation crucial for quantum
 physics. It is well known, strongly compact closed categories for categorical
 quantum mechanics \& information \cite{coecke1,coecke2,coecke3} are special examples
 of {\em dagger ribbon categories} where the braiding is a symmetry (permutation) and
 the twist operator is the identity, and modular tensor categories with the dagger
 operation, responsible for the mathematical description of topological quantum
 computing \cite{flw1,flw2}, are also special cases for {\em dagger ribbon categories}.

 But this proposal will raise a natural question about the roles that the
 braiding and twist play in {\em categorical quantum physics and information}.
 Possible answers have been discussed in the first footnote.
 ``Braiding" suggests that physical objects either fundamental ones at the
 Planck energy scale or quasi-particles of many-body systems are required to
 obey the braiding statistics, while `twist" means that they are either
 string-like (even braine-like), i.e., extended configurations instead
 of point particles, or have an effective (a new internal) degree of freedom
 called the ``twist spin". The latter one can be commented from the similar historical
 story how an electron was found to have a spin quantum number different from
 its known orbital angular momentum quantum number. Therefore, a quasi-particle
 obeying the braiding statistics like an anyon either has the ``twist spin", or
 behaves in a string-like way so that the configuration formed by its motion can
 be denoted by a strip or ribbon in which the existence of the ``twist spin"
 is natural, in addition, this anyon should live with the so called quantum
 dimension specified by ribbon tensor categories.

 As concluding remarks, {\em categorical quantum physics} has been explained
 word by word: ``categorical" by {\em dagger ribbon categories}
 \footnote{The classification of dagger ribbon categories will be a worthwhile
 problem to be considered for both mathematicians and physicists.}; ``quantum" by
 the superposition principle of state  vectors in a Hilbert space as well as the
 complex structure (such as the  imaginary unit $i$) and unitary evolution of
 a state vector; and ``physics" by specific physical topics to be described in the
 framework of categories. Furthermore, it is explicit that the name {\em categorical
 quantum physics and information} survives different sorts of
 interpretations, for example,
 ``categorical" can be related to other categorical structures (even categories over
 categories) in which {\em  dagger ribbon categories} are subcategories or
 special examples so that quantum measurement, classical data, etc., can be treated at
 the same time. Moreover, it is crucial to emphasize again that the proposal for
 {\em categorical quantum physics and information} in the first footnote is
 motivated by the present study in quantum information phenomena and theory, to
 be a solution for the problem how to coordinate a mathematical framework in which
 both categorical quantum mechanics \cite{coecke1, coecke2, coecke3} and topological
 quantum computing \cite{flw1,flw2} are interesting examples.
 Finally, dagger ribbon categories involved in this article are those with
 positive definite forms between two morphisms, i.e., positive dagger ribbon
 categories which are called unitary Hermitian ribbon categories in Turaev's book
 \cite{turaev}.

\section*{Acknowledgements}

 Y. Zhang is indebted to S. Abramsky for helpful discussions and
 email correspondences during the writing of the draft,
 thanks B. Coecke and P. Selinger for stimulating discussions at the
 workshop ``Cats, Kets and Cloisters" (Computing Laboratory,
 Oxford University, July 17-23, 2006),  and thanks
 S.L. Braunstein and R. Griffiths for critical comments on the early
 draft. He is grateful for Y.-S. Wu for stimulating discussions, and
 thanks E.C. Rowell for helpful email correspondences on Hermitian ribbon
 categories.

 The main body of this manuscript had been completed by Y. Zhang
 during his visiting period (with the host, M.-L. Ge) in Chern
 Institute of Mathematics, Nankai University, China, and he is in
 part supported by the seed funding of University of Utah and NSFC
 Grant-10605035. L.H. Kauffman is in part supported by NSF Grants.

\appendix

\section{The extended TL algebra}

The extended TL algebra, i.e., the TL algebra with local unitary
transformations, has been proposed to underlie quantum circuits in
terms of maximally entangled states and local unitary
transformations \cite{yong1,yong2}. The TL algebra $TL_n$ is
generated by identity $Id$ and $n-1$ hermitian projectors $E_i$
satisfying \eqa \label{tl}
 E_i^2 &=&  E_i, \qquad (E_i)^\dag=E_i,\,\,\, i=1,\ldots,n-1, \nonumber\\
 E_i E_{i\pm1} E_i &=& \lambda^{-2} E_i, \qquad E_i E_j=E_j E_i, \,\,\, |i-j|>1,
  \ena
in which $\lambda$ is called the loop parameter.

A representation of the $TL_n(d)$ algebra is obtained in terms of
the maximally entangled state $\omega$, a projector, by defining
idempotents $E_i$ in the way \eq E_i=(Id)^{\otimes (i-1)} \otimes
\omega \otimes (Id)^{\otimes (n-i-1)}, \qquad i=1,\cdots n-1.\en For
example, the $TL_3(d)$ algebra is generated by two idempotents $E_1$
and $E_2$, \eq
 E_1=\omega\otimes Id, \qquad E_2=Id\otimes \omega.
\en In Figure 12, there are diagrammatical representations for
$E_i$, $E_1 E_2$ and $E_1 E_2 E_1 =\frac 1 {d^2} E_1$ with the loop
parameter $d$. $E_1 E_2$ has a normalization factor $\frac 1 d$ from
a vanishing cup and a vanishing cap, and $E_1 E_2 E_1 $ has a factor
$\frac 1 {d^2}$ from two vanishing cups and two vanishing caps.

 \begin{figure}
\begin{center}
\epsfxsize=13cm \epsffile{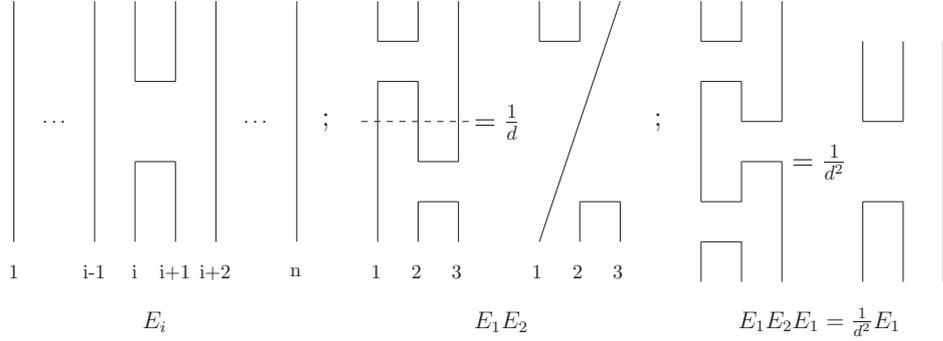} \caption{Teleportation
configuration and TL algebra.} \label{fig12}
\end{center}
\end{figure}

With local unitary transformations $U_n$ of the maximally entangled
state $\omega$, one can set up another representation of the TL
algebra. For example, the $TL_3(d)$ algebra is generated by
 $\tilde{E}_1$ and $\tilde{E}_2$,
 \eq \tilde{E}_1=\omega_n\otimes Id, \qquad \tilde{E}_2=Id \otimes
\omega_n. \en Therefore, the extended TL algebra has diagrammatical
configurations consisting of cups, caps and solid points or small
circles which have been exploited to describe quantum circuits
\cite{yong1,yong2}.

Furthermore, the extended TL algebra is also an interesting
mathematical framework for performing quantum computing at the
diagrammatical level. For example, the CNOT gate, a linear
combination of tensor products of Pauli matrices $\sigma_1$,
$\sigma_2$ and $\sigma_3$,
 \eq
 C= \frac 1 2 (1\!\! 1_2 \otimes 1\!\! 1_2 +1\!\! 1_2\otimes \sigma_1
 +\sigma_3\otimes 1\!\! 1_2  -\sigma_3\otimes \sigma_1) \en which
 satisfies the properties of the CNOT gate, \eq
 C|00\rangle=|00\rangle, \,\, C|01\rangle=|01\rangle,\,\,
 C|10\rangle=|11\rangle,\,\, C|11\rangle=|10\rangle,  \en
 has an extended TL diagrammatical representation Figure 13. Similarly,
 a swap gate (symmetry) can be represented by a linear combination of
 the extended TL configuration, see \cite{yong2} for an example, which
 verifies that the Brauer category \cite{brauer} is a kind of extended
 TL category.

 \begin{figure}
\begin{center}
\epsfxsize=12.5cm \epsffile{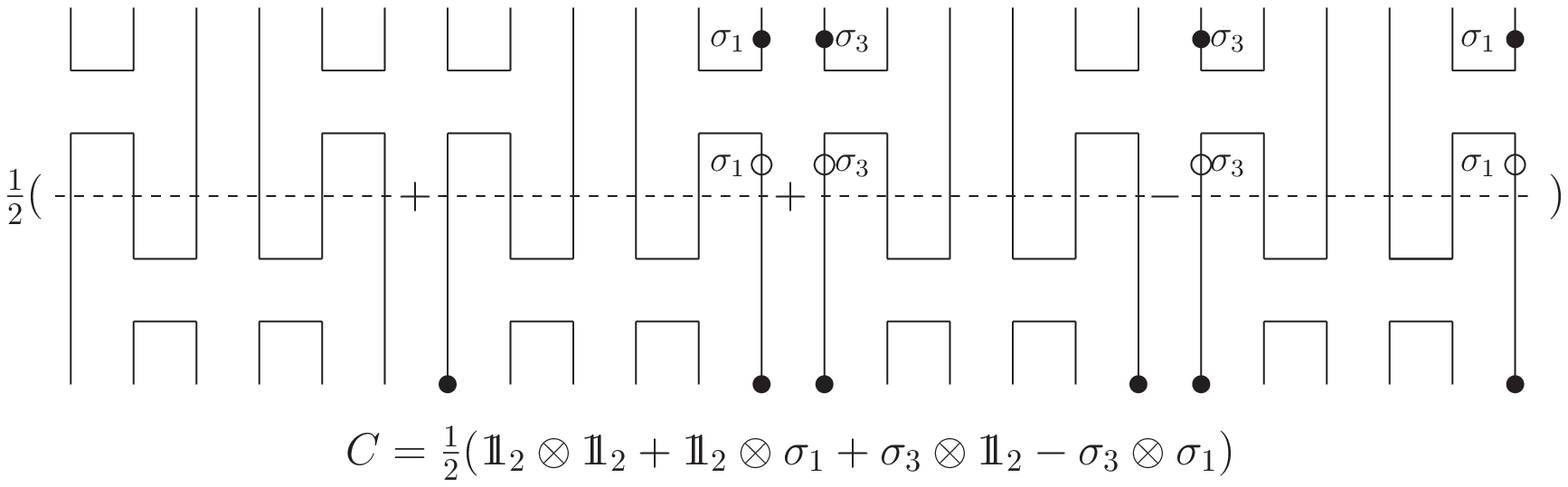} \caption{The CNOT gate in
our diagrammatical approach.} \label{fig13}
\end{center}
\end{figure}

 \section{Definitions of Categories}

Definitions of various kinds of categories used in this paper have
been sketched in the following. They include categories, monoidal
categories (tensor categories), pivotal categories, dagger pivotal
categories, braided monoidal categories, symmetric monoidal
categories, symmetric pivotal categories (compact closed
categories), symmetric dagger pivotal categories (strongly compact
closed categories or 3-tuply categories with duals), ribbon
categories, modular tensor categories and $d$-dimensional cobordism
categories.  References for them are referred to Kassel's textbook
\cite{kassel} and Turaev's book \cite{turaev}.

A {\em category} ${\mathcal C}$ consists of objects $A, B,C,
\cdots$, associative morphisms $f, g,h, \cdots$, \eq  \frac {f:
A\rightarrow B, \,\, g: B\rightarrow C,\,\, }{ g\circ f:
A\rightarrow C,\,\,},\qquad ( h\circ g) \circ f =
 h \circ (g\circ f),  \en and identity morphism  $Id$
 between these objects,
 \eq f\circ {Id_A}=f,\qquad {Id_B}\circ f=f,\qquad
    Id_A: A\rightarrow A,
     \qquad Id_B: B\rightarrow B.
\en A morphism is also called an arrow. The inverse of a morphism
$f$ is denoted by $f^{-1}: B\rightarrow A$. A {\em functor} $Z$
between two categories $\mathcal  C$ and ${\mathcal C}^\prime$ is a
structure-preserving map in the following sense, \eq \frac{f:
A\rightarrow B} {Z(f): Z(A)\rightarrow Z(B)},\qquad Z(f\circ
g)=Z(f)\circ Z(g),\,\,\, Z(Id_A)=Id_{Z(A)} \en

A {\em monoidal category} $({\mathcal C},\otimes, I, \alpha, l, r)$
is a category ${\mathcal C}$ equipped with a tensor product
$\otimes$ on objects and morphisms,
 \eq \frac {f: A\rightarrow B, \qquad
 g: C\rightarrow D} {f \otimes g: A\otimes C \rightarrow B \otimes D }
 \en together with unit object $I$, associative
natural isomorphism $\alpha$, left and right natural unit
isomorphisms $l$ and $r$,  \eq \alpha_{A,B,C}: (A\otimes B)\otimes C
 \stackrel{\cong}{\rightarrow} A\otimes (B\otimes C),\qquad
 l_A: A \stackrel{\cong}{\rightarrow} I\otimes A,\qquad
  r_A: A \stackrel{\cong}{\rightarrow} A\otimes I, \en
where the symbol $\cong$ denotes the isomorphism relation. The
associator $\alpha$ has to satisfy the pentagon equation on the
tensor product $((A\otimes B)\otimes C)\otimes D$,
 \eq (Id_A\otimes \alpha_{B,C,D})\circ (\alpha_{A,B\otimes C,D})
 \circ(\alpha_{A,B,C}\otimes Id_D)
 =(\alpha_{A,B,C\otimes D})\circ ( \alpha_{A\otimes B,C,D} ),
 \en
while the associator $\alpha$, left and right units $l,r$ have to
satisfy the triangle equation on the tensor product $A\otimes
I\otimes B$, \eq
 (Id_A \otimes l_B)\circ (\alpha_{A,I,B}) =( r_A\otimes Id_B).
\en The pentagon and triangle equations are also called coherent
laws. As the associator $\alpha$ is an identity morphism, the
monoidal category is called strict. In the literature, monoidal
categories are also called {\em tensor categories}.

A {\em pivotal category} is a  monoidal category with the left
duality $()^\ast$ which assigns a dual object $A^\ast$ to any object
$A$ and imposes a unit morphism $\eta_A$ and a counit morphism
$\epsilon_A$ given by \eq \epsilon_A:
 A^\ast \otimes A \rightarrow I, \qquad \eta_A: I \rightarrow
  A\otimes A^\ast \en satisfying the triangular identities, \eqa
 l^{-1}_{A^\ast}\circ (\epsilon_A\otimes Id_{A^\ast})\circ
  \alpha^{-1}_{A^\ast,A,A^\ast}
 \circ(Id_{A^\ast}\otimes \eta_A)\circ r_{A^\ast} &=& Id_{A^\ast},
 \nonumber\\
 r_A^{-1}\circ ( Id_A \otimes \epsilon_A ) \circ \alpha_{A,A^\ast,A} \circ
  ( \eta_A \otimes Id_A)\circ l_A &=& Id_A,
 \ena which are also called the rigid conditions. The left duality
 $()^\ast$ acts on objects in the manner,
 \eq
A^{\ast\ast} \cong A, \qquad (A\otimes B)^\ast=B^\ast\otimes A^\ast,
\qquad I^\ast=I
 \en
and it is a contravariant involutive functor, \eq
 (Id_A)^\ast=Id_{A^\ast}, \qquad (g\circ f)^\ast=f^\ast\circ g^\ast, \qquad
 f^{\ast\ast}=f  \en where the transpose $f^\ast:
B^\ast\rightarrow A^\ast$ of the morphism $f:A\rightarrow B$ is
defined by \eq
 f^\ast = ( \epsilon_B \otimes Id_{A^\ast})\circ
 (Id_{B^\ast} \otimes f \otimes Id_{A^\ast})\circ (Id_{B^\ast}\otimes \eta_A
 ).
\en

A category can be equipped with an involutive, identity-on-objects,
contravariant functor $()^\dag: {\mathcal C}\rightarrow {\mathcal
C}^{op}$ which defines the adjoint $f^\dag: B\rightarrow A$ of a
morphism $f: A\rightarrow B$ satisfying \eq
 A^\dag \cong A, \qquad (Id_A)^\dag=Id_A, \qquad
 (g\circ f)^\dag =f^\dag \circ g^\dag, \qquad f^{\dag\dag}=f.
\en This adjoint functor $()^\dag$ can coherently preserve the
tensor product structure in a monoidal category, \eq
 (A\otimes B)^\dag \cong A\otimes B, \qquad
  (f\otimes g)^\dag \cong f^\dag \otimes g^\dag,
\en and the adjoints of the associator $\alpha$, left and right
units $l, r$ are given by \eq
 \alpha^\dag=\alpha^{-1},\qquad l^\dag=l^{-1},\qquad r^\dag=r^{-1}.
\en A {\em dagger pivotal category} is called for a monoidal
category with the left duality $()^\ast$ and $()^\dag$ functor,
which allows $\eta_A^\dag=\epsilon_{A^\ast}$
 and $\epsilon_{A}^\dag=\eta_{A^\ast}$ and leads to  a covariant functor
 $()_\ast$ denoting the complex conjugation $f_\ast$ of the morphism $f$,
 \eq
 \frac {f: A\rightarrow B} {f_\ast: A^\ast \rightarrow B^\ast},
 \qquad f \mapsto f^{\ast\dag}; \qquad
 (f^\ast)_\ast=f^\dag=(f_\ast)^\ast.
 \en

A {\em braided monoidal category} is a monoidal category with a
braiding, $\sigma_{A,B}: A\otimes B \rightarrow B\otimes A$, a
natural isomorphism satisfying \eq
 (g\otimes f)\circ \sigma_{A,B} = \sigma_{B^\prime, A^\prime} \circ
 (f\otimes g), \qquad f: A\rightarrow B^\prime,\,\,\, g: B\rightarrow
 A^\prime
\en for all morphisms $f$ and $g$, and also satisfying the hexagonal
equations, \eqa (Id_B\otimes \sigma_{A,C} )\circ \alpha_{B,A,C}
\circ (\sigma_{A,B}\otimes Id_C ) &=&
 \alpha_{B,C,A}\circ\sigma_{A,B\otimes C}\circ  \alpha_{A,B,C},
 \nonumber\\
 (\sigma_{A,C}\otimes Id_B ) \circ \alpha^{-1}_{A,C,B}
 \circ (Id_A\otimes \sigma_{B,C}) &=& \alpha^{-1}_{C,A,B}\circ\sigma_{A\otimes
 B,C}\circ \alpha^{-1}_{A,B,C}.
\ena As the braiding $\sigma_{A,B}$ forms a representation of the
symmetrical group, i.e., satisfying $\sigma_{A,B}\circ
\sigma_{B,A}=Id_{A\otimes B}$ for all objects $A$ and $B$, this
category is called the {\em symmetric monoidal category}.

In the literature, symmetric pivotal categories  are called  {\em
compact closed categories} or {\em rigid symmetric monoidal
categories}, and {\em dagger symmetric  pivotal categories}
\cite{selinger1} are called {\em strongly compact closed categories}
by Abramsky and Coecke \cite{coecke1,coecke3} or {\em 3-tuply
categories with duals}  by Baez and Dolan \cite{baez1}. The {\em
compact closed category} was introduced by Kelly \cite{kelly1} in
the 1970's, in particular an important paper by Kelley and Laplaza
on coherence in compact closed categories \cite{kelly2}.

A twist $\theta_A$ in the braided monoidal  categories is a natural
isomorphism $\theta_A: A\rightarrow A$ satisfying \eq
 \theta_{A\otimes B}=(\theta_A \otimes \theta_B) \circ
 \sigma_{B,A}\circ \sigma_{A,B}, \qquad
 \theta_{A^\ast}=(\theta_A)^\ast
 \en
for all objects $A, B$ in the category. {\em Ribbon categories} are
braided monoidal categories with the left duality $()^\ast$ and a
twist $\theta_A$. {\em Modular tensor categories} are a kind of
special ribbon categories, and they have simple objects $X_1, X_2,
\cdots X_n$ with the fusion rule $X_i \otimes X_j \cong
 \oplus_{i,j=1}^n C_{ij}^k X_k$, fusion coefficients $C_{ij}^{k}$
to be natural numbers (or zero) and an invertible symmetric
$s$-matrix by its matrix entries $s_{ij}=Tr(\sigma_{X_i,X_j}\circ
\sigma_{X_j,X_i} )$. In the first footnote, {\em dagger ribbon
categories}. i.e., ribbon categories with the dagger operation
$()^\dag$ coherently preserving the ribbon structure,
$\sigma_{A,B}^\dag =\sigma_{A,B}^{-1}$,
$\theta^\dag_A=\theta^{-1}_A$ and \eq
 \eta_A^\dag=\epsilon_A\circ \sigma_{A,A^\ast}\circ (\theta_A \otimes
 Id_{A^\ast}),\quad \epsilon_A^\dag=(Id_{A^\ast}\otimes
 \theta_A^{-1})\circ\sigma_{A^\ast,A}^{-1}\circ\eta_A
\en have been proposed to be a underlying mathematical framework for
{\em categorical quantum physics and information}. With the
convention by Abramsky and Coecke for strongly compact closed
categories, {\em dagger ribbon categories} can be also called
strongly ribbon categories, see Paquette's unpublished thesis
\footnote{In this unpublished reference (informed by Coecke in his
feedback to Zhang on the first web version of the present
manuscript), dagger ribbon categories are involved but all of those
key proposals in the first footnote and last section are not
realized and made.} \cite{paquette}.

Since Hilbert space is an object with positive definite forms
between two linearly bounded operators, dagger ribbon categories in
the proposal for {\em categorical quantum physics and information}
have to be equipped with positive definite forms between two
morphisms, i.e., positive dagger ribbon categories also called
unitary Hermitian ribbon categories in Turaev's book \cite{turaev}.
But for dagger ribbon categories with some specific fusion rules,
one can not achieve both positive definite forms and unitary
braidings due to compatibility conditions between them, see
interesting examples in Rowell's papers \cite{rowell1, rowell2}. The
manuscript \cite{yong5} to be an extension of the proposal in the
first footnote and last section will completely exploit Turaev's
notations on Hermitian ribbon categories.

Besides categories introduced as above, the $d$-dimensional {\em
cobordism category} {\bf Cob[d]} to define TQFT  by Atiyah
\cite{atiyah}, has as its objects smooth manifolds of dimension $d$,
and as its morphisms, smooth manifolds $M^{d+1}$ of dimension $d+1$
with a partition of the boundary, $\partial M^{d+1}$, into two
collections of $d$-manifolds that we denote by $L(M^{d+1})$ and
$R(M^{d+1})$. We regard $M^{d+1}$ as a morphism from $L(M^{d+1})$ to
$R(M^{d+1})$, $ M^{d+1}: L(M^{d+1}) \Longrightarrow R(M^{d+1})$.
These categories {\bf Cob[d]} are highly significant for quantum
physics  \& information, especially {\bf Cob[0]} is directly related
to Dirac notations of quantum mechanics and to the Brauer category
\cite{brauer} or TL category \cite{lieb}.


\begin{thebibliography}{99}

  \bibitem{nielsen}
  M. Nielsen and I. Chuang, {\it Quantum Computation and Quantum Information}
 (Cambridge University Press, 1999).

\bibitem{coecke1} S. Abramsky, and B. Coecke,   {\it A Categorical
Semantics of Quantum Protocols}. In:  Proceedings of the 19th Annual
IEEE Symposium on Logic in Computer Science  (LiCS`04), IEEE
Computer Science Press. Arxiv:quant-ph/0402130.

\bibitem{coecke2} B. Coecke,  {\it Kindergarten Quantum Mechanic--Lecture Notes}.
   In: {\em Quantum Theory: Reconstructions of the Foundations III}, pp. 81-98,
    A. Khrennikov,  American Institute of Physics Press.
    Arxiv: quant-ph/0510032.

\bibitem{coecke3} S. Abramsky, {\it Temperley-Lieb Algebra: from Knot Theory to
Logic and Computation via Quantum Mechanics}, To appear in
Mathematics of Quantum Computation and Quantum Technology, G. Chen,
L. Kauffman and S. Lomonaco, eds., Taylor and Francis, pages
523--566, 2007.

\bibitem{kauffman1}
 L.H. Kauffman and S.J. Lomonaco Jr.,  {\it Braiding Operators are
 Universal Quantum Gates},  New J. Phys. {\bf 6} (2004) 134.
 Arxiv: quant-ph/0401090.

 \bibitem{kauffman2}
 L.H. Kauffman, {\it Teleportation Topology}.
 Opt. Spectrosc. {\bf 9} (2005) 227-232. Arxiv: quan-ph/0407224.

\bibitem{griffths} R.B. Griffiths, S. Wu, L. Yu and S. M. Cohen,
  {\it Atemporal Diagrams for Quantum Circuits},
  Phys. Rev. {\bf A 73} (2006) 052309. Arxiv: quant-ph/0507215.


  \bibitem{yong1} Y. Zhang, {\it Teleportation, Braid Group and TL Algebra}.
   J. Phys. {\bf A:} Math. Gen. {\bf 39} (2006) 11599-11622. Arxiv:
   quant-ph/0610148.

  \bibitem{yong2} Y. Zhang, {\it Algebraic Structures Underlying Quantum Information Protocols}.
   Arxiv: quant-ph/0601050.

\bibitem{lieb} H.N.V. Temperley and E.H. Lieb, {\it Relations between the `Percolation' and
  `Colouring' Problem and Other Graph-Theoretical Problems Associated with Regular Planar Lattices:
  Some Exact Results for the `Percolation' Problem}, Proc. Roy. Soc. A {\bf 322} (1971) 251-280.


 \bibitem{ZKW} Y. Zhang, L.H. Kauffman and R.F. Werner, {\it Permutation
 and its Partial Transpose},  Arxiv: quant-ph/0606005. Accepted
     by {\em International Journal of Quantum Information} for
     publication.

  \bibitem{ZKW1} Y. Zhang, L.H. Kauffman and M.L. Ge, {\it Virtual Extension
  of Temperley--Lieb Algebra}. Arxiv: math-ph/0610052.

 \bibitem{brauer}  R. Brauer, {\it On Algebras Which Are Connected With the
 Semisimple Continuous Groups}, Ann. of Math. {\bf 38} (1937) 857-872.

 \bibitem{kauffman3} L.H. Kauffman and S.J. Lomonaco Jr.,
 {\it $q$-Deformed Spin Networks, Knot Polynomials and Anyonic
 Topological Quantum Computation}, Arxiv: quant-ph/0606114.

 \bibitem{bennett} C.H. Bennett, G. Brassard, C. Crepeau, R.
  Jozsa, A. Peres and W. K. Wootters, {\it  Teleporting an Unknown Quantum State
  via Dual Classical and Einstein-Podolsky-Rosen Channels}, Phys. Rev. Lett. {\bf 70}
  (1993) 1895-1899.

 \bibitem{vaidman} L. Vaidman, {\it Teleportation of Quantum States},
 Phys. Rev. {\bf A} 49 (1994) 1473-1475.

 \bibitem{eriz}  N. Erez, {\it Teleportation from a Projection Operator Point of
 View.}  Arxiv: quant-ph/0510130.

\bibitem{braunstein} S.L. Braunstein, G.M. D'Ariano, G.J. Milburn
 and M.F. Sacchi, {\it Universal Teleportation with a Twist}, Phys.
 Rev. Let. {\bf 84} (2000) 3486--3489.

 \bibitem{ekert}
M. \.{Z}ukowski, A. Zeilinger, M.A. Horne and A.K. Ekert,  {\it
`Event-Ready-Detectors' Bell Experiment via Entanglement Swapping}.
Phys. Rev. Let. {\bf 71} (1993) 4287--4290.

\bibitem{atiyah} M.F. Atiyah, {\it The Geometry and Physics of
 Knots} (Cambridge University Press, 1990).

  \bibitem{wu} F.Y. Wu, {\it Knot Theory and Statistical Mechanics},
  Rev. Mod. Phys. {\bf 64} (1992) 1099-1131.

 \bibitem {kauffman4} L.H. Kauffman, {\it Knots and Physics}
  (World Scientific Publishers, 2002).

 \bibitem{jones} V.F.R. Jones, {\it Heck Algebra Representations of Braid Groups
  and Link Polynomials},  Ann. of Math. {\bf 126} (1987) 335-388.

 \bibitem{werner} R. F. Werner,  {\it All Teleportation and Dense Coding
  Schemes}, J. Phys. A {\bf 35} (2001) 7081--7094.  Arxiv: quant-ph/0003070.

 \bibitem{joyal} A. Joyal and R. Street, {\it The Geometry of Tensor Calculus
 I}, Adv. in Math. {\bf 88} (1991) 55-112.

 \bibitem{kelly1} G.M. Kelly, {\it Many-Variable Functional Calculus I}.
 Spinger Lecture Notes in Mathematics {\bf 281} (1972) 66-105.

 \bibitem{kassel} C. Kassel, {\it Quantum Groups} (Spring-Verlag New York, 1995).

 \bibitem{turaev} V.G. Turaev, {\it Quantum Invariants of Knots and 3-Manifolds}
   (de Gruyter, 1994).

\bibitem{baez1} J. Baez and J. Dolan, {\it High-dimensional Algebra and Topological
 Quantum Field Theory}, JMP {\bf 36} 6703-6105. Arxiv:
 q-alg/9503002.

\bibitem{baez2} J. Baez,  {\it Quantum Quandaries: a
Category-Theoretic Perspective}. In: S.~French et al. (Eds.) {\it
Structural Foundations of Quantum Gravity}, Oxford University Press.
Arxiv:quant-ph/0404040.

 \bibitem{selinger1} P. Selinger,  {\it Dagger Compact Closed
Categories and Completely Positive Maps}.  Electronic Notes in
Theoretical Computer Science (special issue: Proceedings of the 3rd
International Workshop on Quantum Programming Languages).

 \bibitem{levin} M. Levin and X.-G. Wen, {\it String-net Condensation: A Physical
 Mechanism for Topological Phases},  Phys. Rev. B {\bf 71}(2005)
 045110.

 \bibitem{yong3} Y. Zhang, L.H. Kauffman and M.L. Ge,
 {\it Universal Quantum Gate, Yang--Baxterization and Hamiltonian}.
  Int. J. Quant. Inform., Vol. 3, {\bf 4} (2005) 669-678. Arxiv: quant-ph/0412095.

 \bibitem{yong4} Y. Zhang, L.H. Kauffman and M.L. Ge,
 {\it Yang--Baxterizations, Universal Quantum Gates and
  Hamiltonians}.  Quant. Inf. Proc. {\bf 4} (2005) 159-197.
  Arxiv: quant-ph/0502015.

\bibitem{witten} E. Witten, {\it Quantum Field Theories and the
 Jones Polynomial}, Comm. Math. Phys. {\bf 121} (1989) 351-399.

 \bibitem{flw1} M. Freedman, M. Larsen, and Z. Wang, {\it A Modular Functor
 Which is Universal for Quantum Computation}. Arxiv: quant-ph/0001108.

 \bibitem{flw2}
M. H. Freedman,  A. Kitaev, Z. Wang, {\it Simulation of Topological
Field Theories by Quantum Computers}, {Commun. Math. Phys.}, {\bf
227}, 587-603 (2002). Arxiv: quant-ph/0001071.

 \bibitem{segal} G. Segal, {\it Two-dimensional Conformal Field Theories and Modular Functors},
 IXth. International Congress on Mathematical Physics, Swansea, 1988, 22¨C37.

 \bibitem{smolin} C.Rovelli and L. Smolin, {\it Spin Networks and Quantum
 Gravity}, Phys. Rev. D {\bf 52} (1995) 5743-5759.

 \bibitem{kelly2} G.M. Kelly and M.L. Laplaza, {\it Coherence for Compact Closed
 Categories}, J. Pure Appl. Algebra {\bf 19} (1980) 193-213.

 \bibitem{paquette} E.O. Paquette, {\it A Categorical Semantics for Topological
    Quantum Computation}, Master's thesis, University of Ottawa (2004).

 \bibitem{rowell1} E.C. Rowell, {\it  On a Family of Non-Unitarizable Ribbon
 Categories}, Math. Z.  {\bf 250} no. 4 (2005) 745-774.

  \bibitem{rowell2} E.C. Rowell, {\it From Quantum Groups to Unitary Modular Tensor
  Categories}, to appear in Contemp. Math. (Conference Proceedings).

 \bibitem{yong5} Y. Zhang,  {\it Categorical Quantum Physics and
    Information}, in preparation.


 \end{thebibliography}
\end{document}